# Domain Wall Acceleration by Ultrafast Field Application: An Ab Initio-Based Molecular Dynamics Study

*Ruben Khachaturyan,* Aris Dimou, and Anna Grünebohm*

**Optimizing ferroelectrics for contemporary high-frequency applications asks for the fundamental understanding of ferroelectric switching and domain wall (DW) motion in ultrafast field pulses while the microscopic understanding of the latter is so far incomplete. To close this gap in knowledge, ab initio-based molecular dynamics simulations are utilized to analyze the dynamics of 180° DWs in the prototypical ferroelectric material $BaTiO_3$. How ultrafast field application initially excites the dipoles in the system and how they relax to their steady state via transient negative capacitance are discussed. Excitingly, a giant boost of the DW velocity related to the nonequilibrium switching of local dipoles acting as nucleation centers for the wall movement is found. This boost may allow to tune the local ferroelectric switching rate by the shape of an applied field pulse.**

## 1. Introduction

Switchable polarization makes ferroelectrics indispensable in contemporary electronics.[1–7] In particular, ultrafast switching[8,9] becomes increasingly important for memory devices,[10,11] neuromorphic computing,[12,13] and telecommunication.[14] To optimize the switching process, one needs to fundamentally understand and control the formation and movement of domain walls (DWs), that is, interfaces between regions with the different polarization directions.[15,16]

Although investigations on DW movement have a long history and it is settled that field-driven DW movement is mediated by nucleation and growth,[17–22] it turned out that it is important to revisit this concept with modern high-resolution experiments and atomistic simulations.[16] For example, the shape and size of critical nuclei could be identified only recently[15,23–25] and a field-dependent transition between DW propagation and homogeneous nucleation in the domains has been predicted.[26]

In particular, if it comes to ultrafast pulses, a complete fundamental understanding is so far missing and the THz response of ferroelectric materials is an active field of research.[13,27,28] First reports on fast field changes hint to exotic transient effects with potential impact on the DW dynamics,[29–32] including a giant electrocaloric response[33] and the so-called transient negative capacitance (NC).[34,35] Although being counterintuitive at first glance, the NC can be understood by dipoles that switch against the field due to local depolarization fields[35–43] which could be related to temporarily uncompensated bound charges.[44] Already Landauer[45] predicted that NC and nucleation and growth of clusters with reverse polarization are closely interlinked and thus a large impact on the DW motion is likely. Recently, Parsonnet et al.[31] observed deceleration of the field-induced DW motion with time and related this to the NC but a microscopic understanding is so far missing.

In this letter, we present a first systematic microscopic study on the coupling between local dipole dynamics, the transient NC, and the motion of 180° DWs in the tetragonal phase of $BaTiO_3$ under ultrafast field applications. Excitingly, we find a temporary giant speed-up of DWs with high potential for fast and efficient devices based on short field pulses.

## 2. Experimental Section

We opt for ab initio-based molecular dynamic simulations using the *feram* code[46] in combination with the effective Hamiltonian by Zhong et al.[47,48] and the parametrization by Nishimatsu et al.[49] The collective atomic displacements were coarse grained to the most relevant degrees of freedom per unit cell, the acoustic displacement $\vec{w}$ and the soft mode $\vec{u}$, which corresponded to the local dipole moment $\vec{p} = Z^*\vec{u}$, with $Z^*$ the effective Born charge. This Hamiltonian was given as

$$\begin{aligned} H^{\text{eff}} = {} & V^{\text{self}}(\{\vec{u}\}) + V^{\text{dpl}}(\{\vec{u}\}) + V^{\text{short}}(\{\vec{u}\}) \\ & + V^{\text{elas}}(\eta_1, \ldots, \eta_6, \{\vec{w}\})) \\ & + V^{\text{coup}}(\{\vec{u}\}, \{\vec{w}\}, \eta_1, \cdots, \eta_6) \\ & - Z^* \sum_i \vec{E}_{\text{ext}} \cdot \vec{u}_i + \frac{M^*_{\text{dipole}}}{2} \sum_{i,\alpha} \dot{\vec{u}}^2_{\alpha,i} \end{aligned} \tag{1}$$

R. Khachaturyan, A. Dimou, A. Grünebohm
Interdisciplinary Centre for Advanced Materials Simulation (ICAMS) and Center for Interface-Dominated High Performance Materials (ZGH)
Ruhr-University Bochum
Universitätsstr 150, 44801 Bochum, Germany
E-mail: ruben.khachaturyan@rub.de

The ORCID identification number(s) for the author(s) of this article can be found under https://doi.org/10.1002/pssr.202200038.

where $V^{self}(\{\vec{u}\})$, $V^{dpl}(\{\vec{u}\})$, and $V^{short}(\{\vec{u}\})$ are the self-energy, the long-range dipole–dipole interaction, and the short-range interactions of the local soft modes; the elastic energy $V^{elas}$ depends on $\{\vec{w}\}$ and the homogeneous strain tensor $\eta_i$ given in the Voigt notation, and $V^{coup}$ includes the couplings between local soft mode and strain. The last two terms are the coupling to the external field $\vec{E}_{ext}$ and the kinetic energy of the local mode, with $M^*_{dipole}$ being the mass of the soft mode.

Hereby only the time evolution of the local mode is explicitly treated in molecular dynamic simulations while the strain is internally optimized. This makes the method extremely efficient and allows us to use a simulation box of 164 × 48 × 48 u.c. (65 × 19 × 19 nm), making the method appealing to study the microscopic domain structure evolution in ferroelectrics. After it has been used successfully to model functional properties and domain structures of $BaTiO_3$,[50–54] we now present a first systematic study of the wall dynamics for ultrafast field application.

Finite-size effects and thermal noise are small in these large systems. We furthermore validate our results by independent simulation runs with different random initialization of dipoles. The error bar is an order of magnitude smaller than the calculated velocities; for more details see Figure F2, Supporting Information.

As we use periodic boundary conditions, we actually model a periodic array of 180° DWs with equidistant spacing. However, due to the large cell, the walls are initially separated by 82 u.c and their distance does not go below 10 u.c. in the time interval of interest. Thus, the interaction between neighboring walls can be neglected and the discussion focuses on the wall initially located at $x_n = 1$, as shown in **Figure 1**.

The Nosé-Poincaré thermostat[55] and a time step of 1 fs are utilized to equilibrate the system in the tetragonal phase.[56] We focus on one temperature being in the stability range of the tetragonal phase as found by our model: 260 K which is 40 K below the cubic-to-tetragonal transition temperature and 125 K above the tetragonal-to-orthorhombic transition temperature. Thereby, we initialize a multidomain structure by applying local fields of ±10 kV mm$^{-1}$ which are removed in four steps (±7, ±4, ±1, and 0 kV mm$^{-1}$). After each step, we equilibrate the system for 30 ps. Finally, a homogeneous electric field is applied along $z$ which increases linearly during the ramping time $t_{ramp}$ and is then kept fixed for the rest of the simulation time, that is, we use a field pulse given as $E_{ext}(t) = E_{ext} \cdot \left(\frac{t}{t_{ramp}} \cdot H(t_{ramp} - t) + H(t - t_{ramp})\right)$, where $H(t_{ramp})$ is the Heaviside function. While we monitor the DW dynamics for ramping times up to 100 ps and a field up to 60 kV mm$^{-1}$, our discussion of the underlying microscopic mechanisms focuses on the limiting case of instantaneous field application ($t_{ramp} = 0$ ps) and a field magnitude of 40 kV mm$^{-1}$. We note that the absolute values of the applied fields cannot be compared directly to experimental values as molecular dynamics simulations on idealized bulk materials overestimate critical field strengths for polarization switching.[43,57]

The DW velocity ($v_{DW}$) is given by the shift of the DW center $x_0$ with time. To track the time evolution of $x_0$, we fit the mean-polarization per $x$-layer $\langle p_z \rangle_x$ every 1 ps via

$$\langle p_z \rangle_x = p_0 \cdot \tanh\left[\frac{x - x_0}{d_{DW}}\right] + \varepsilon(x) \quad (2)$$

where the first term corresponded to the polarization profile of the DW without an external field[58] with $p_0$ and $d_{DW}$ being the saturation polarization and the width of the wall, respectively. The second term $\varepsilon(x)$ is a simple linear expression to account for polarization response on the applied field within domains, including linear dielectric response and transient NC.

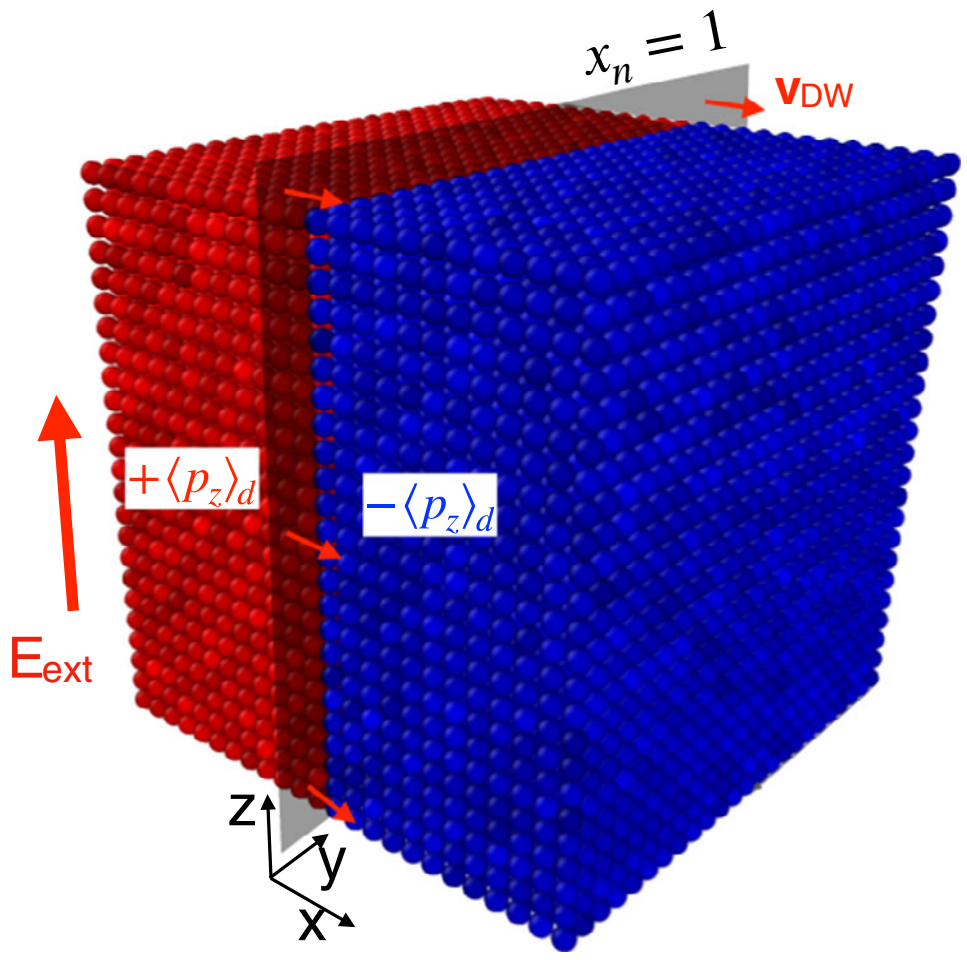

**Figure 1.** Excerpt of the simulation cell visualized with Ovito.[67] Each sphere represents one unit of $BaTiO_3$ color encoded by the sign of the polarization along $z$. The 180° DW initialized in plane $x_n = 1$ separates layers with mean polarization of $\pm\langle p_z \rangle_x$. After the electric field $E_{ext}$ is applied, the wall moves along $x$ with velocity $v_{DW}$.

## 3. Results and Discussion

While domains with polarization along ±$z$ are equivalent in the initial state, applying a field along $z$ favors the positive polarization direction and gradually switches the system to the single-domain state. This switching is possible by the growth of the positive domain (wall motion) or the homogeneous switching in the whole material with the latter facing a larger energy barrier. In agreement to literature,[26] our systematic screening of different field strengths thus discloses three field regimes. We find a first critical field strength of 5 kV mm$^{-1}$ for the onset of DW motion. Second, between 5 and 40 kV mm$^{-1}$, the switching is mainly driven by DW motion and its velocity increases with $E_{ext}$. Finally for $E_{ext} > 40$ kV mm$^{-1}$, we observe DW motion and superimposed nucleation and growth of positive clusters in the negative domain. Note that the absolute values of the critical field strengths depend on temperature and cannot be directly compared with experimental values due to the well-known overestimation of critical field strengths in MD simulations.

Excitingly, the initial DW speed increases by a factor of 4 if we reduce the ramping time $t_{ramp}$ from 100 ps to 0 ps, see **Figure 2**. Furthermore, $v_{DW}$ depends strongly on field strength and time, as shown in **Figure 3**a. One may distinguish three time intervals: in the time interval (I, pink area), we find an unexpected field-dependent boost of the velocity reaching its maximum around 1 ps. During the transient interval (II, blue area), the velocity decays drastically with time, and in the steady-state (III, gray area), we find the expected increase in the velocity with the field strength (0.5 km s$^{-1}$ under 20 kV mm$^{-1}$ and 1 km s$^{-1}$ under 40 kV mm$^{-1}$)[59,60] and a superimposed small linear decrease in wall velocity with time.

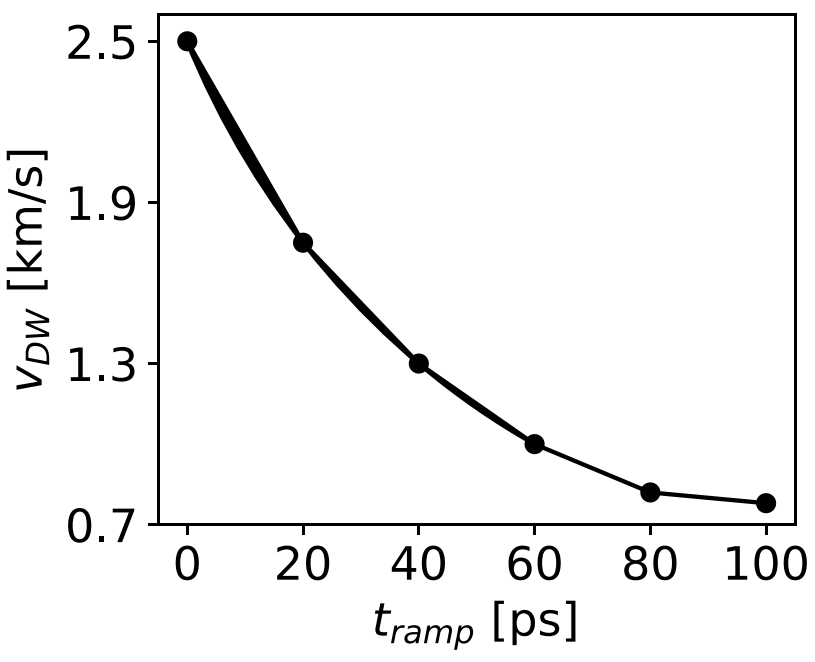

**Figure 2.** Enhancement of the wall velocity $v_{DW}$ with increasing rate of the field change ranging from $t_{ramp} = 100$ ps corresponding to the linear increase in the field over 100 ps to $t_{ramp} = 0$ being the instantaneous application of the field ($E_{ext} = 40$ kV mm$^{-1}$, at $t = 1$ ps).

It turns out that all these trends are related to the time evolution of the polarization in the negative domain, see Figure 3. In all layers, $x_n$, both velocity and layer-wise polarization increase abruptly in the time interval (I), followed by their fast decay in the time interval (II), which corresponds to the transient NC, and finally a linear reduction in the time interval (III).

The linear reduction of $\langle p_z \rangle_{x_n}$ can be understood by the coupling between positive and negative domains. Although, one would intuitively expect that the magnitudes of polarization and strain in separate positive and negative domains increase and decrease in the field, respectively, both domains are coupled, and thus the growing positive domain successively enhances strain and polarization magnitude in the entire system, see increase in polarization magnitudes in both domains between 13 ps and 40 ps in **Figure 4**a. The larger the field, the larger is the increase in strain and polarization magnitude. It has been discussed in the literature that the energy barrier for polarization switching scales with polarization magnitude.[58] Consequently, one also has to expect a reduction of the DW velocity with time and an increasing slowing down with increasing field strength, fully in line with our observations.

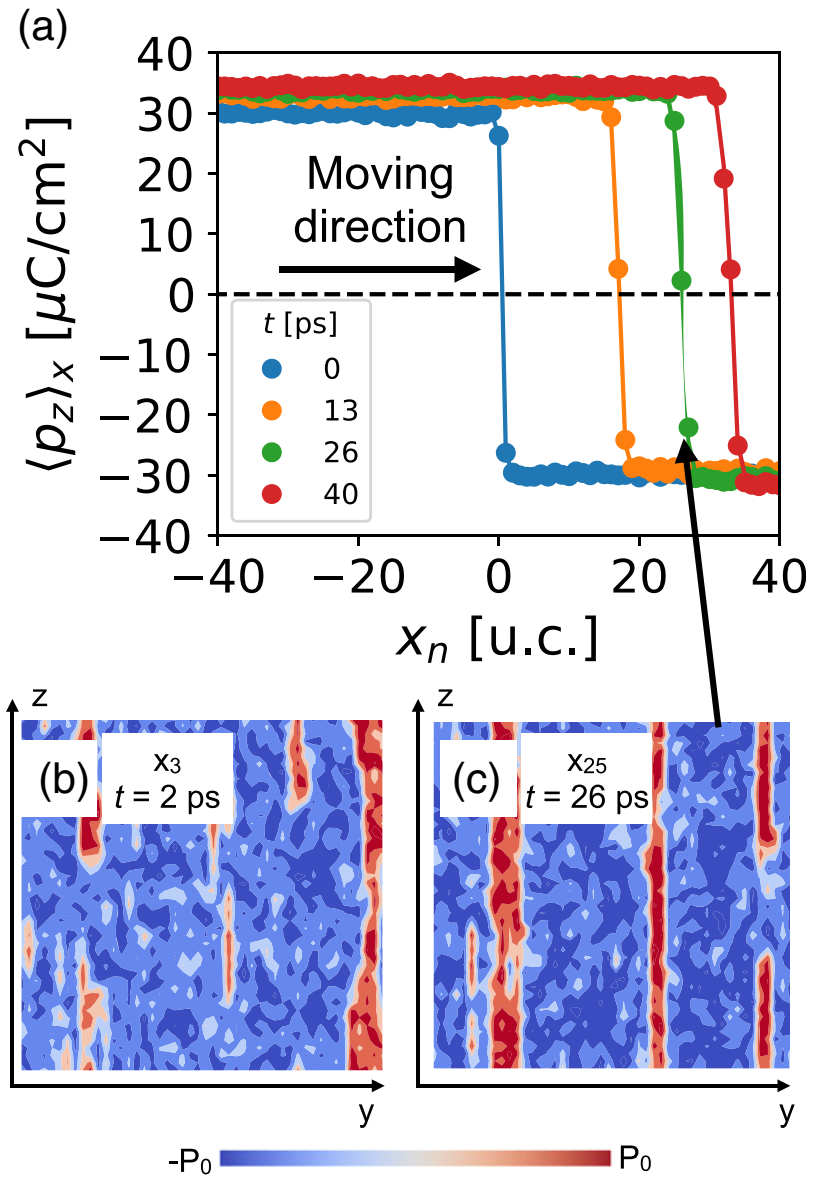

**Figure 4.** a) Field-induced shift of the DW after instantaneous field application ($E_{ext} = 40$ kV mm$^{-1}$) shown by the change of layer-resolved polarization $\langle p_z \rangle_x$ with time (color of dots). Lines show the fit Equation (2) used to track the time evolution of the center and width of the wall. After field application, the wall broadens by the formation and growth of clusters with reversed polarization in front of the moving interface, as shown by representative snapshots of layers $x_n$. b) A larger amount of smaller clusters form in nonequilibrium regime (time interval (II)) on the layer $x_3$ compared with c) propagation in a steady state as shown for $x_{25}$ at 26 ps with larger needle-like clusters for the same amount of reversed dipoles. Colors encode $p_z$.

This linear time dependency of polarization magnitude and DW velocity can neither explain the drastic boost of the wall velocity at the beginning of the simulations, nor its fast decay in the time interval (II). The latter observations can be understood by the following detailed analysis of the microscopic dipole distribution, only.

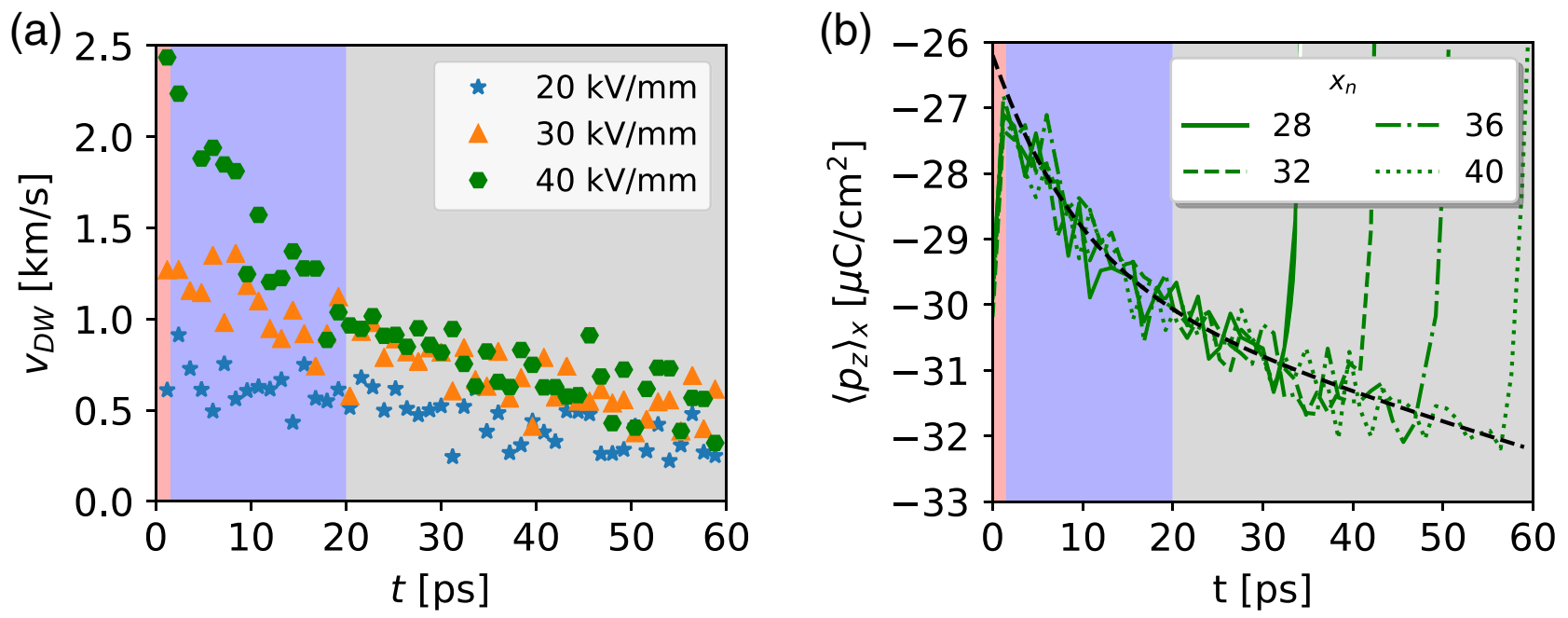

**Figure 3.** a) The initial boost of the DW velocity $v_{DW}$ and its time-dependent decay in comparison with b) the time evolution of the polarization in representative layers $\langle p_z \rangle_{x_n}$ under 40 kV mm$^{-1}$. Three different time regimes are highlighted by background colors: (I, pink area) initial velocity boost and dipole switching, (II, blue area) transient state with slowing down of the wall and transient NC, and (III, gray area) steady state with small linear reduction of $v_{DW}$ and increase in $|\langle p_z \rangle_x|$ with time. Different colors of the data points display the used field strengths and the line types distinguish different layers $x_n$. The overall decay of the velocity with time corresponds to the sum of a linear and an exponential function (black line).

The starting point for our discussion is the time evolution of the layer-resolved polarization profile $\langle p_z \rangle_x$ across the wall, as shown in Figure 4a. Initially, the DW is centered at $x_n = 1$ and we determine a width of 1.2 u.c. in agreement to previous reports.[58,61,62] After field application, the wall broadens and layer(s) with reduced polarization magnitude form. This reduction is mainly due to the formation and growth of clusters with reversed polarization, as shown in Figure 4b,c. One has to distinguish between the steady state of the system and the non-equilibrium initial state after fast field application. First many small reversed clusters appear on the wall surface, see subfigure (b). This configuration will be discussed in more detail below. Later, in the steady state, mainly needle-like larger clusters are present; see subfigure (c). The formation of such larger clusters is in agreement with previous reports and can be related to the smaller activation energy for the growth of clusters compared with the energy barrier to nucleate a new cluster, as discussed in the work. (Ref. [23]).[23] In turn, once a cluster of reversed dipoles has nucleated it quickly grows along the polarization direction resulting in the needle-like clusters which expand in the plane until the full layer has been reversed. If the switching barrier was unchanged in the steady state, the constant velocity would be observed, but in our case, domains' strain modifies the energy profile; thus, slight linear velocity decrease is present.

To understand the drastic changes of the wall propagation after field applications, it is important to note that also apart from the DWs, the local polarization is never perfectly homogeneous at finite temperatures and thermal fluctuations result in the temporary reversal of 0.4% of the dipoles against the surrounding polarization already without external field. Although the applied field favors the parallel alignment of dipoles, the applied field of 40 kV cm$^{-1}$ and thermal fluctuations are too small to overcome the energy barrier for homogeneous polarization switching; see **Figure 5**. Fast field changes however drive the system out of thermodynamic equilibrium[63,64] and as soon as the field is applied, up to 4% of the dipoles in the negative domain spontaneously align with the field direction. The larger the applied field, the more the dipoles have enough energy to overcome the energy barrier to switch, and we find an increase in the amount of reversed dipoles up to 2% for 20 kV mm$^{-1}$, 2.6% for 30 kV mm$^{-1}$, and up to 4% for 40 kV mm$^{-1}$ by initial switching, see also Figure F1a, Supporting Information. The importance of the nonequilibrium character of this switching is underlined by the fact that the number of initially switched dipoles also depends nonlinearly on $t_{\text{ramp}}$, that is, 2.4%, 1.5%, and 1.25% of dipoles switch for $t_{\text{ramp}}$ 20, 60, and 100 ps, respectively.

For the understanding of how these initially switched dipoles may temporally boost the DW velocity, it is important to discuss their properties in more detail. Remarkably, the non-equilibrium switching does not only increase the number of reversed dipoles but also reduces the spatial correlation of dipoles, as can be seen from snapshots of the same layer after 1 ps in Figure 5a and after 25 ps in Figure 4c. The initially switched dipoles in Figure 5a form many small clusters distributed over the whole negative domain in contrast to the large clusters found on the moving wall in its steady state in Figure 4c. This change of the spatial correlation can be quantified by the Pearson correlation coefficient[65] (see Formula E1) for that layer which increases from $R = 0.55$ (initial switching) to $R = 0.91$ (steady state).

The difference in the distribution between initially switched dipoles and growing nuclei on the moving wall also results in different energies of the system. If single dipoles or small clusters flip relative to their surrounding, the resulting polarization gradients induce a finite density of bound charges $\rho_z = -(\partial p_z/\partial z)$, as shown I Figure 5b. According to Landau–Ginzburg–Devonshire's theory, this charge density induces an energy penalty of $U_z = g_z \rho^2$, with the material-specific constant $g_z$.[66] In agreement with previous reports on the microscopic structure of reversed clusters in ferroelectrics,[23,25] we find that the system reduces the polarization gradients by the reduction of local polarization and diffuses interfaces of the clusters; see Figure 5a. However, a finite energy penalty remains. Using $g_z = 0.38 \cdot 10^{-11} \text{m}^3/\text{F}$ of $BaTiO_3$ as obtained by ab initio calculations in the study by Shin et al.,[23] we find that an increase in the polarization by $1\mu\text{C cm}^{-2}$ by initial switching results in an energy penalty of $U_z = 0.17\text{meV/u.c.}$ This value is remarkably high, considering the fact that the same increase in polarization on the moving wall (after half of the dipoles in a layer has been switched) only accounts for $U_z = 0.1\text{meV/u.c.}$

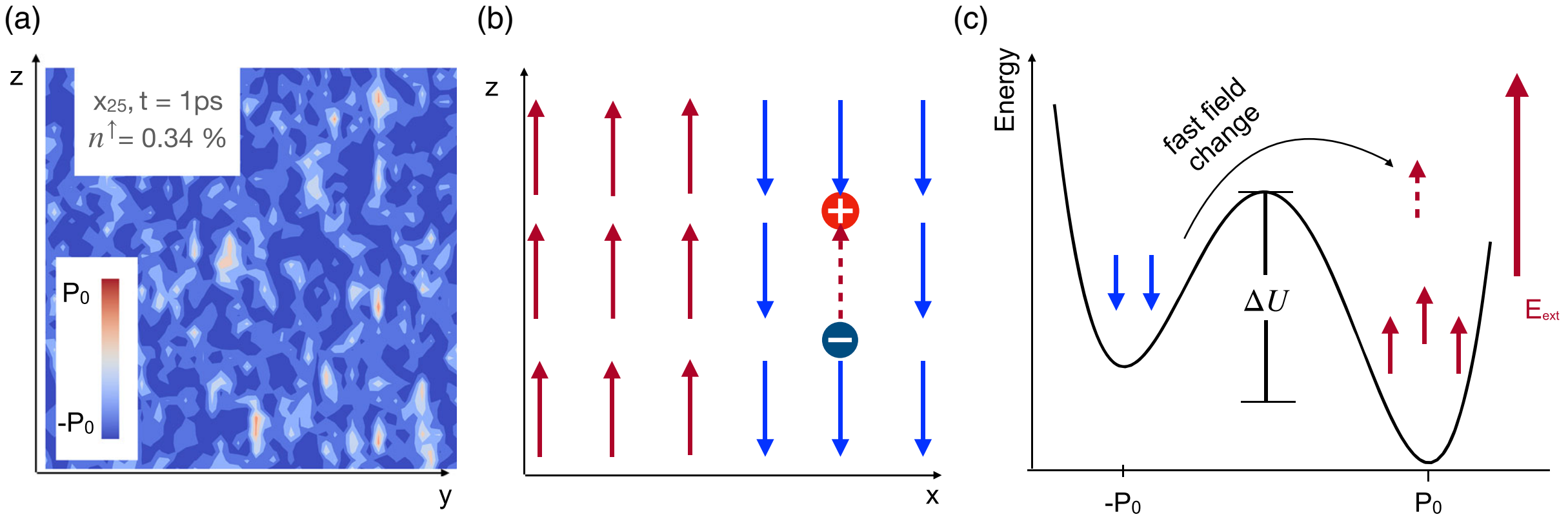

**Figure 5.** Distribution of the initially switched dipoles and its impact on the energy of the system. a) After initial switching (1 ps), many small clusters in the whole negative domain are aligned with the field direction as shown for representative layer $x_{25}$. b) Sketch of the system with initial switching in xz-plane: If a dipole is switched in the negative domain (dashed arrow), it results in a polarization gradient, induces a local bound charge density, and thus an energy penalty. c) Sketch of the energy landscape: Although the applied field $E_{\text{ext}}$ lowers the energy of the parallel polarization direction, the energy barrier $\Delta U$ for homogeneous switching is too large to be overcome by fluctuations in thermal equilibrium. Fast field changes bring the system out of equilibrium and several dipoles flip to the unfavorable configuration (dashed arrow shown in (b)).

How can these small switched clusters with high energy boost the DW movement? As discussed earlier, the rate-limiting process for the DW motion in the steady state is the nucleation of reversed clusters at the wall interface. For this, local dipoles have to switched by thermal fluctuations. Analogous to point defects,[21] the dipoles initially switch by the fast field pulse, lower the energy to flip their surroundings due to the existing polarization gradients, and may thus act as nucleation centers and accelerate the DW motion. Indeed the switching on the wall in time intervals ($I$) is not related to the growth of a few large clusters as in the steady state. Rather many smaller clusters start to grow at the initially switched dipoles, see Figure 4b, accelerating the switching of the full layer to 3.6 ps only (2.5 km $s^{-1}$). In contrast, the same process without the initial nucleation centers takes 6 ps, corresponding to a wall velocity of 1 km $s^{-1}$, only. We note that the initial switching may also nucleate clusters of critical size in the domain, resulting in the homogeneous polarization switching reported in the study by Boddu et al.[26] For 40 kV $mm^{-1}$, we however find a small probability to form a critical nucleus apart from the wall.

Nonequilibrium initial switching and the resulting polarization gradients do not only boost the nucleation process for polarization switching and wall movement, but the higher energy of the initial dipole configuration is also the reason for the transient NC in the time interval (II): The dipoles in the center of the negative domain switch back to the more favorable homogeneous polarization antiparallel to the applied field, see the time evolution of $n^{\uparrow}$ in Figure F1b, Supporting Information, and thus the polarization "increases" against the applied field. Obviously, the back-switching of dipoles also reduces the number of potential nucleation centers and thus the transient NC is the source of the slowing down of the DW with time.

## 4. Conclusion

We have used molecular dynamics simulations to study the impact of the field application rate on ferroelectric switching and field-induced DW motion. We find that fast field changes drive the system out of thermal equilibrium and thereby give rise to two interlinked exciting properties: 1) a large temporary boost of the DW velocity and 2) transient NC. This finding is of particular interest for ferroelectric memory devices.

Our detailed microscopic analysis reveals that both effects are related to the switching of local dipoles with low spatial correlation. On the one hand, these dipoles can act as nucleation centers for polarization switching in front of the travelling wall and thus boost the wall velocity. On the other hand, this inhomogeneous distribution of dipoles induces unfavorable local bound charges and thus the dipoles switch back with time, which is the origin of the transient NC in the system.

Although we restricted our study to the qualitative trends for linear field ramping and 180° walls in $BaTiO_3$, a similar time dependency of the DW motion is likely for other shapes of field pulses, materials, and DWs as soon as local dipoles with low correlations are switched out of their equilibrium positions and the transient NC may serve as a fingerprint for potential realizations of the effect.

## Supporting Information

Supporting Information is available from the Wiley Online Library or from the author.

## Acknowledgements

The authors acknowledge financial support guaranteed by the Deutsche Forschungsgemeinschaft (DFG) via the Emmy Noether group GR4792/2 and thank Dr. Madhura Marathe for fruitful discussion.

Open Access funding enabled and organized by Projekt DEAL.

## Conflict of Interest

The authors declare no conflict of interest.

## Data Availability Statement

The data that support the findings of this study are available from the corresponding author upon reasonable request.

## Keywords

ferroelectric domain walls, functional oxides, ultrafast switching